\documentstyle[aps,epsf,amsmath,amsfonts,amssymb,amsbsy]{revtex}
\begin{document}
\title{Lifshitz points in blends of AB and BC diblock copolymers}
\author{P.~D. Olmsted$^1$\footnote {{\tt p.d.olmsted@leeds.ac.uk}} and
  I.~W. Hamley$^2$\footnote{{\tt i.w.hamley@chem.leeds.ac.uk}}}
\address{ $^1$Department of Physics and Astronomy and Polymer IRC, and
  $^2$School of
  Chemistry; University of Leeds, Leeds LS2 9JT, United Kingdom} %

\maketitle

\begin{abstract}
  We consider micro- and macro- phase separation in blends of $AB$ and
  $BC$ diblock copolymers. We show that, depending on architecture, a
  number of phase diagram topologies are possible. Microphase
  separation or macrophase separation can occur, and there are a
  variety of possible Lifshitz points. Because of the rich parameter
  space, Lifshitz points of multiple order are possible. We
  demonstrate Lifshitz points of first and second order, and argue
  that, in principle, up to 5th order Lifshitz points are possible.
\end{abstract}
%
%
%
%

\section{Introduction}
The phase behaviour of block copolymer melts is remarkably rich.  In a
blend of homopolymers only macrophase separation (with wavenumber
$q_{\ast}=0$) occurs.  Macrophase separation in a block copolymer melt
is prevented by the chemical connectivity of the constituent blocks,
which leads to microphase separated structures with $q_{\ast} \neq 0$,
typically corresponding to structural periods $L\simeq10\!-\!100\,{\rm
  nm}$ \cite{hamley,batesglenn90}. In a blend containing a block
copolymer melt and one or more molten homopolymers, microphase
separation of the block copolymer can compete with macrophase
separation of the homopolymers at low temperatures \cite{hamley}.

In a binary blend of a block copolymer and a homopolymer, the
homopolymer swells the microphase separated structure formed by the
copolymer, if the homopolymer chain length is less than or equal to
that of the corresponding block \cite{hamley,hasegawa}. On the other
hand, macrophase separation can occur for homopolymer chains longer
than the corresponding block. In a ternary blend, block copolymer
added to a blend of homopolymers acts as a compatibilizer to prevent
macrophase separation or reduce the lengthscale associated with the
macrophase separated structure \cite{hamley,roe87}. A similar
interplay between micro- and macro- phase separation has recently been
explored experimentally for $AB/AB$ diblock copolymer blends by
Hashimoto and coworkers \cite{hashimoto94}. Recently, self-consistent
field theory has been applied to examine the phase behavior of binary
homopolymer/copolymer blends \cite{whitmore85,Mats95c,janert96},
blends of two homopolymers with block copolymer
\cite{whitmore85b,janert97} and binary blends of block copolymers
\cite{MatsBate95b,shi95}. Particularly interesting critical phenomena
have been predicted for certain blends of copolymer with one or two
homopolymers. The latter case was first studied using Landau mean
field theory, employing the random phase approximation (RPA)
\cite{brosetta90,holyst92}. In addition to lines of critical points
corresponding to macrophase separation or microphase separation, mean
field theory predicts that Lifshitz points can occur at the boundary
between disordered, uniformly ordered and periodically ordered phases
\cite{brosetta90,holyst92}. The wavenumber for microphase separation
approaches zero continuously as the Lifshitz point is approached
\cite{lifshitz}. A Lifshitz point was first observed experimentally in
the phase diagram for blends of two polyolefin homopolymers and the
corresponding diblock via small-angle neutron scattering by Bates
\textit{et al.}  \cite{Bate+95}. However, subsequent work showed that
composition fluctuations destroy the mean field Lifshitz point and a
microemulsion phase becomes stable \cite{bates97}. Mean field theory
can then be used to locate the region of microemulsion stability via
the virtual Lifshitz point.

In contrast to these studies of copolymer/homopolymer blends and
blends of AB diblocks, we are unaware of any experimental work on
blends of an AB diblock with a BC diblock. This letter presents some
predictions for these systems which should stimulate future
experimental work. We employ the RPA, first applied to AB diblocks by
Leibler \cite{leibler80}, to locate spinodal points for macro- or
micro- phase separation, and to compute the wavenumber and eigenvector
of the unstable mode.  This approach is expected to be valid for long,
weakly segregated, chains. Generalization of the approach outlined
here to allow for composition fluctuations and finite chain length
should be straightforward, using methods developed for pure block
copolymer melts \cite{glennhelfand87,olmsted94b}. A theory for micelle
formation in blends of strongly segregated AB and BC diblocks has
recently appeared \cite{borovinskii98}, however we are unaware of any
previous work on the weak segregation regime of these systems.

\section{Model}
Let $\phi$ be the volume fraction of the $AB$ diblock; $f$ and $\beta
f$ the fractions of the $A$ and $C$ components in the $AB$ and $BC$
copolymers, respectively; and $N$ and $\alpha N$ the respective
monomer numbers. For simplicity we assume equal monomer volume and
statistical segment length for all species. We work in terms of a
vector of fluctuations $\boldsymbol{\psi}$,

\begin{equation}
\label{eq:psi}
\boldsymbol{\psi} = \{\psi_{\rm A},\psi_{\rm B}, \psi_{\rm C}\},
\end{equation}
where $\psi_{\alpha}$ is the deviation of the volume fraction of
species $\alpha$ from its mean value. It is straightforward to
calculate the correlation functions
\begin{equation}
G_{\alpha\beta}(q)=\langle\psi_{\alpha}(q)\psi_{\beta}(-q)\rangle
\end{equation}
using the RPA \cite{leibler80}, including three Flory $\chi$
parameters $\chi_{\rm AB}, \chi_{\rm AC},$ and $\chi_{\rm BC}$. It is
convenient to define the basis set
\begin{equation}
\label{eq:basis}
\boldsymbol{e}_0= \sqrt{\tfrac13}\left\{1,1,1\right\}, \boldsymbol{e}_1=
\sqrt{\tfrac23}\left\{\tfrac12,-1,\tfrac12\right\}, \boldsymbol{e}_2=
\sqrt{\tfrac12}\left\{1,0,-1\right\}, 
\end{equation}
where $\boldsymbol{\psi}\!\cdot\!\boldsymbol{e}_0$ is a volume
changing fluctuation and $\boldsymbol{\psi}\!\cdot\!\boldsymbol{e}_1$
and $\boldsymbol{\psi}\!\cdot\!\boldsymbol{e}_2$ are physical
fluctuations in an incompressible system. The fluctuation
$\boldsymbol{\psi}\!\cdot\!\boldsymbol{e}_1=\,\sqrt{3/2}(\psi_{\rm A}
+ \psi_{\rm C})$ corresponds to separating the $A$ and $C$ blocks from
the $B$ block, and is primarily a microphase separation mode, since it
is prohibited at $q=0$ by chain connectivity. The other
mode,$\boldsymbol{\psi}\!\cdot\!\boldsymbol{e}_2 =
\sqrt{1/2}\,(\psi_{\rm A}-\psi_{\rm C})$, corresponds to demixing the
$A$ and $C$ blocks, and in the limit $q\rightarrow 0$ corresponds to
demixing the blend.  Hence we term this a macrophase separation mode.
A general fluctuation at $q\neq 0$ is an admixture of these two modes,
while only mode $\boldsymbol{e}_2$ is present for $q=0$.

The spinodal is given by the determinant of the $2\times 2$ matrix of
$G_{\alpha\beta}(q)$ in the incompressible
$\{\boldsymbol{e}_1,\boldsymbol{e}_2\}$ subspace, \begin{equation}
\label{eq:gamma}
\Gamma(q) = G_{11}(q)G_{22}(q) - G_{12}(q)^2, \end{equation}
where
$G_{ab}(q)=\boldsymbol{e}_a\!\cdot\!\boldsymbol{G}\!\cdot\!\boldsymbol{e}_b$.
$\Gamma(q)$ is a product of the fluctuation eigenvalues. These
eigenvalues have minima at $q=0$ (macrophase separation) or
$q_{\ast}\neq 0$ (microphase separation). The spinodal
point is given by that eigenmode whose eigenvalue first vanishes upon
reducing the temperature. For $q=0$ this eigenmode is
$\boldsymbol{e}_2$, while otherwise it is an admixture of
$\boldsymbol{e}_1$ and $\boldsymbol{e}_2$. 
The small-$q$ expansion of $\Gamma$ has the form
\begin{equation}
\label{eq:gamma0}
\Gamma(q) = \frac{a_0 + a_1 q^2 + a_2 q^4 + a_3 q^6 +  \ldots}{b_1  q^2}.
\end{equation}

To parametrize the problem we let $\chi\equiv\chi_{AB}N,
r_{AC}\equiv\chi_{AC}/\chi_{AB},$ and
$r_{BC}\equiv\chi_{BC}/\chi_{AB}$.  The phase diagram may now be
calculated in the $\chi-\phi$ plane, with
$r_{AC},r_{BC},f,\beta,\alpha$ as independent material parameters.
Obviously the system is far richer (and more complicated) than that of
simple diblocks. Rather than systematically calculating phase
diagrams, we first discuss the nature of macro- and micro-phase
separation, and then examine the character of the possible Lifshitz
points.

\section{Microphase vs. Macrophase separation}
In the $AB$/$AB$ limit ($\chi_{AC}=0,\chi_{AB}=\chi_{BC}$) macrophase
separation cannot occur; while for large enough $\chi_{AC}$ macrophase
separation is possible.  The nature of the unstable modes can be seen
by examining the eigenvalues $\lambda_1(q)$ and $\lambda_2(q)$ of the
fluctuation matrix (in the $2$ dimensional incompressible subspace).
{\begin{figure}[!tbh] \epsfxsize=5.5truein \centerline{\epsfbox[22 40
      731 270]{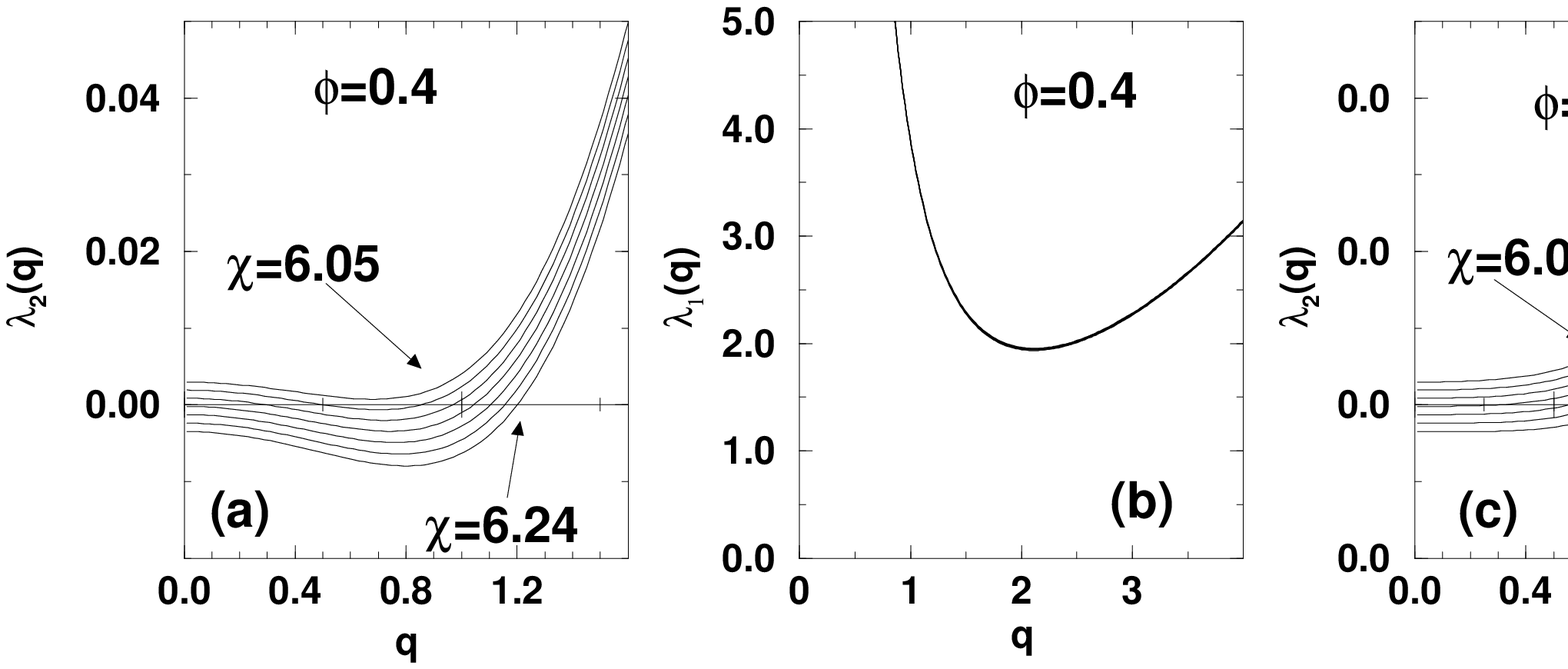}}
\caption{Fluctuation eigenvalues as a function of wavevector  (units
  of $R_g^{-1}$, where $R_g$ is the radius of gyration) for
  $f=0.17,\beta=1,\alpha=1,r_{AC}=0.49,r_{BC}=2.9$, for $\phi=0.4$ and
  $\phi=0.6$ and a range of $\chi$ values. Variations of $\lambda_1$
  with $\chi$ are shown, but not visible (b).}
\label{fig:eigen}
\end{figure}}

Typical results are shown in Fig.~\ref{fig:eigen} for a blend with $f=
0.17,\alpha=\beta=1.0,r_{AC}=0.49,r_{BC}=2.9$, for compositions
$\phi=0.4$ and $\phi=0.6$.  Microphase endpoints for this system occur
at $\phi$ = 0.546 and $\phi$=0.706. Note that one eigenvalue diverges
at $q=0$, and the other is finite. We term these the
\textit{microphase} and \textit{macrophase} modes, respectively.  In
the limit $q\rightarrow 0$, the microphase mode corresponds to
$\boldsymbol{e}_1$ and the macrophase mode to $\boldsymbol{e}_2$,
while at finite $q$ these modes are (orthogonal) linear combinations
of $\boldsymbol{e}_1$ and $\boldsymbol{e}_2$.  Eigenvalues for blends
on either side of the low $\phi$ microphase endpoint are shown in
Figs.~\ref{fig:eigen}a,b and~\ref{fig:eigen}c respectively. At $\phi =
0.4$ a microphase separation transition spinodal is located at
$\chi=6.063$, at which point the local minimum in $\lambda_2(q)$
becomes negative at finite $q_{\ast}$.  The microphase mode shows a
minimum at finite $q$, but remains positive.  The instability of the
macrophase mode can be easily understood, since an $A-B$ homopolymer
melt requires $\chi N\sim 2$ for macrophase separation, and the
corresponding $A-B$ diblock melt requires $\chi N\simeq 10.5$ for
microphase separation.  Hence pure microphase separation is more
costly, and if the system can take advantage of some macrophase
separation (\textit{i.e.} including some component of the eigenvector
$\boldsymbol{e}_2$), it will do so.
\begin{figure}[!tbh]
  \epsfxsize=5.5truein \centerline{\epsfbox[88 280 740
    500]{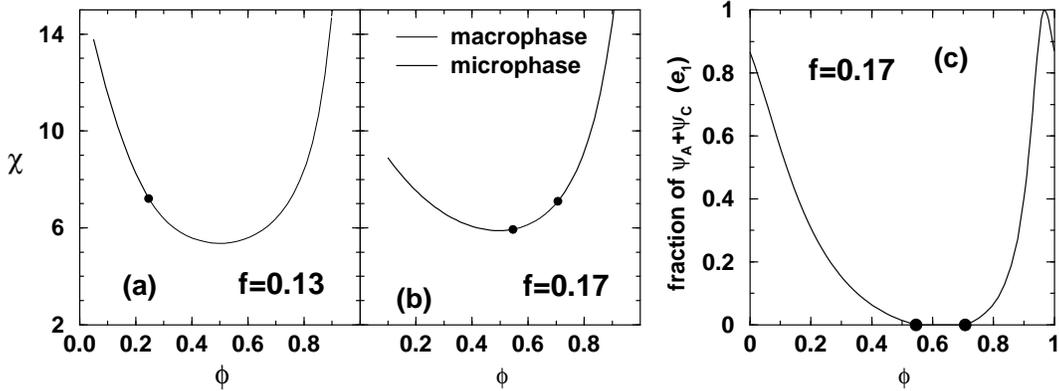}}
\caption{Spinodal diagrams for $\alpha=1,\beta=1$, for $f=0.13$ (a) and
  $f=0.17$ (b), for $\beta=1,\alpha=1,r_{AC}=0.49,r_{BC}=2.9$. Thick
  lines denote microphase spinodals, thin lines denote macrophase
  (liquid-liquid) spinodals, and the filled circles are the microphase
  endpoints. (c) shows the contribution of the microphase separation
  mode $\boldsymbol{e}_1=\sim\psi_A+\psi_C$ along the microphase
  separation lines for $f=0.17$.}
\label{fig:phase}
\end{figure}

Spinodal diagrams are shown in Fig.~\ref{fig:phase}a-b. Since the two
diblocks are identical in architecture and molecular weight, the phase
behaviour results solely from the chemical differences between $A$ and
$C$, through the $\chi$ parameters.  Lowering the the temperature
induces an instability to either macrophase or microphase separation,
depending on copolymer asymmetry and blend composition.  For diblocks
with $f = 0.13$ the disordered phase is unstable to macrophase
separation for $\phi$ near $0.5$, and to microphase separation for
blends with $\phi\lesssim 0.25$ (Fig.~\ref{fig:phase}).  The asymmetry
about $\phi = 0.5$ is due to the distinct temperature dependence of
the three $\chi$ parameters.  Generally the bimodal associated with
the macrophase spinodal ``buries'' the microphase endpoint and we
expect, with increasing $\chi$, macrophase-macrophase coexistence,
macrophase-microphase coexistence, and microphase-microphase
coexistence.  As the copolymers become more symmetric, the region of
macrophase separation narrows, and the criticial point for macrophase
separation coincides with the microphase endpoint at a copolymer
volume fraction $f_L\simeq 0.17$ at a first order Lifshitz point
(Fig.~\ref{fig:phase}b).

Fig.~\ref{fig:phase}c shows the portion of the eigenmode for the
microphase instability which is in fact the microphase eigenmode
$\boldsymbol{e}_1$, along the lines of microphase transitions for
$f=0.17$. At the Lifshitz point (and the other microphase endpoint)
there is an infinitesimal amount of $\boldsymbol{e}_1$, and the
majority of the instability is in the macrophase mode,
$\boldsymbol{e}_2$. As the pure system is approached (either $\phi=0$
or $\phi=1$) the fraction of $\boldsymbol{e}_1$ increases but,
interestingly, does not approach $1$. This is due to the chemical
asymmetry between $A$ and $C$.

\section{Lifshitz Points}
As with a homopolymer blend, the critical composition $\phi_c$ is
given by $\partial a_0/\partial\phi=0$, yielding
$\phi_c=\sqrt{\alpha}/(1+\sqrt{\alpha})$ \cite{pgdgpolymer}. At
$\phi_c$ the critical point $\chi_c$ for macrophase separation is
given by $a_0=0$. If $a_1>0$ macrophase separation occurs directly
from the disordered state; while for $a_1<0$ microphase separation at
finite wavenumber $q_{\ast}$ occurs directly from the disordered state
(and the macrophase separation spinodal is ``buried'').  The limit
$q_{\ast}=0$ defines a point at which the line of microphase
separation transitions meets the spinodal for macrophase separation,
determined by $a_0=a_1=0$.  By tuning the material parameters we can
easily find a first order Lifshitz point, where $a_0=a_1=0$ at the
critical point, $\phi_c$; and a second order Lifshitz point, at which
$a_0=a_1=a_2=0$ at $\phi_c$ \cite{lifshitz}.  In principle, one may
tune the material parameters further to find third ($a_3=0$), fourth
($a_4=0$), and fifth ($a_5=0$) order Lifshitz points. For example, for
fixed $r_{AB},r_{BC}$ and $\beta$ a second order Lifshitz point can be
found by adjusting $\alpha, f$, and $\chi$. A third order Lifshitz
point can, in principle, then be found by adjusting $\beta$ so that
$a_3=0$; and $r_{AB}$ and $r_{BC}$ could then be adjusted to find
fourth and fifth order Lifshitz points (with $a_4=0$ and $a_5=0$
respectively). This is quite a large parameter space, and we have
succeeded only in finding first and second order Lifshitz points.
\begin{figure}[!tbh]
  \epsfxsize=5.5truein \centerline{\epsfbox[70 30 730
    340]{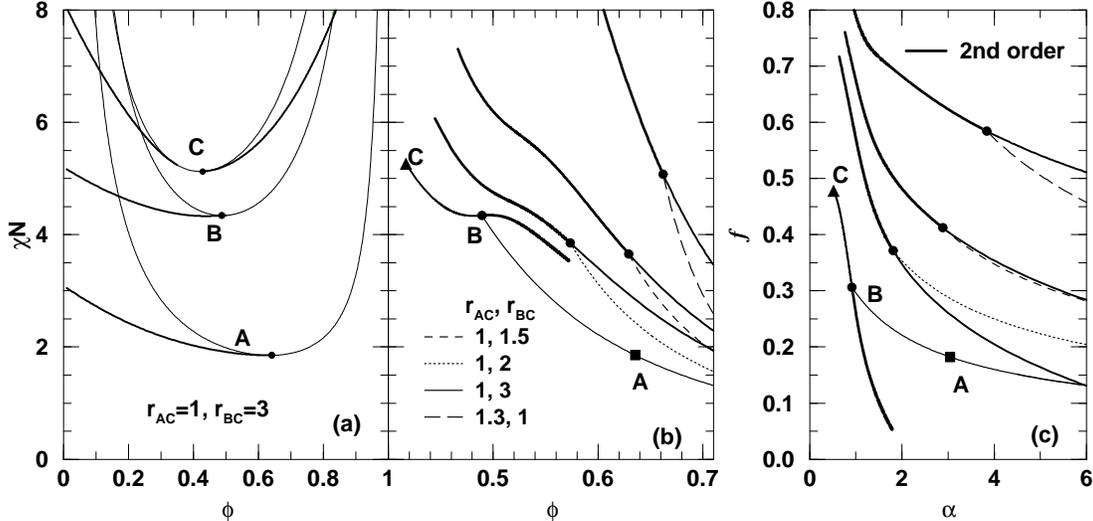}}
\caption{(a): Microphase (thick lines) and macrophase (thin lines) 
  spinodals for points {\bf A} ($\blacksquare$), {\bf B}, and {\bf C}
  $(\blacktriangle)$ in (b) and (c), for $r_{AC}=1, r_{BC}=3$. First
  order Lifshitz points are denoted by $\bullet$.  (b) and (c): Lines
  of Lifshitz points for various $r_{AB},r_{BC}$. Thin lines are first order
  Lifshitz points for $\beta=1$, which end on lines of second order
  Lifshitz points (thick lines) at $\bullet$'s.  Along the second
  order lines $\beta\neq 1$, except for the intersection with the
  first order lines. The ranges of the second order lines are
  $\beta\in(0.590,1.30) [r_{AC}=1,r_{BC}=1.5], \beta\in(0.592,2.32)
  [r_{AC}=1,r_{BC}=2], \beta\in(0.709,5.00) [r_{AC}=1,r_{BC}=3],
  \beta\in(0.64, 1.08)$ $[r_{AC}=1.3,r_{BC}=1]$, where low $\beta$ is
  to the left and high $\beta$ to the right in (b) and (c).}
\label{fig:lif}
\end{figure}

Fig.~\ref{fig:lif} shows lines of Lifshitz points calculated for
various parameters (b,c), and representative spinodal diagrams (a). We
stress that the binodals for macrophase separation, as well as various
microphase structures, will considerably complicate these diagrams.
Nonetheless, the Lifshitz points (\emph{e.g.}  Fig.~\ref{fig:lif}a)
{\bf A} and {\bf C} are the lowest-$\chi$ features in their phase
diagrams, and should be accessible directly from the disordered state.
The Lifshitz lines are shown both in the $\chi\!-\!\phi$ plane
(indicating where in the phase diagram to look), as well as in the
$f\!-\!\alpha$ plane, indicating the trajectory in architecture space.
The first order Lifshitz lines for $\beta=1$ end, at small $\alpha$,
on a second order Lifshitz line which traces out a trajectory in
$\alpha\!-\!\beta\!-\!f$ space. The projections of these lines onto
the $f\!-\!\alpha$ plane are shown as thick lines in
Fig.~\ref{fig:lif}b,c.  The second order lines end at small $\alpha$
(and $\beta$) where a stable root no longer exists; at this point
(such as {\bf C}) the coefficient $a_3$ approaches zero, although our
numerics cannot find a stable solution with $a_0=a_1=a_2=a_3=0$ (which
would signify a third order Lifshitz point). The nature of the
spinodal diagram for {\bf C} suggests that the macrophase separation
window could indeed vanish at third order Lifshitz point for certain
values of the parameters. The higher-order Lifshitz behavior is
indicative of more than one length scale competing for stability, as
would be expected for diblocks which each have a preferred
lengthscale. For large $\alpha$ (and $\beta$), the second order
Lifshitz lines remain stable and do not end.

\section{Summary}
We have examined some aspects of phase separation in $AB/BC$ diblock
copolymer blends.  Both macro- and micro-phase separation can occur,
and microphase separation is a combination of the fundamental
macrophase and microphase eigenmodes. We have demonstrated the
possiblity of Lifshitz points of first and second order, and our
calculations (limited at present by numerical precision) suggest that
Lifshitz points of up to 5th order are, in principle, possible. This
is the first prediction of which we are aware for higher order
Lifshitz points. Clearly, these calculations are illustrative of a
rich phase behaviour which can be mapped by varying architecture and
the three chi parameters. Future work should address the nature of the
ordered microphase-separated phases, and allow for composition
fluctuations. In particular, particularly strong fluctuations are
expected near higher order Lifshitz points [the upper critical
dimension for a $k$th order Lifshitz point is $d_c=4(1+k)$].

\acknowledgments
IWH acknowledges stimulating discussions with collegues in the
EU-TMR programme on ``Complex Architectures in Diblock Copolymer-Based
Polymer Systems''.

\end{document}